%% file: cbpowell_eps_hep_2011_proceedings.tex
\documentclass{PoS}

\usepackage{amsbsy}

\title{Quarkonium Production at STAR}

\ShortTitle{Quarkonium Production at STAR}

\author{\speaker{Christopher Powell (for the STAR Collaboration)}%
\\
 Lawrence Berkeley National Laboratory, 1 Cyclotron Rd, Berkeley, California 94720, USA\\
 University of Cape Town, Lovers Walk Street, Cape Town, South Africa\\
E-mail: \email{CBPowell@lbl.gov}}

\abstract{

We report the results on heavy quarkonium production in $\pp$ and Au+Au collisions at midrapidity via 
the dielectron decay channel at $\sqrt{s_{NN}} = 200$ GeV from the STAR experiment at RHIC. 
Results on the J/$\psi$ $\pt$ spectra in $\pp$ collisions are 
presented for $2 < \pt < 10 \ \gevc$.
The $B$-meson feed-down contribution to the inclusive $\jpsi$ yield has been obtained using the $\jpsi$-hadron azimuthal angular correlation in $\pp$ collisions, 
and is found to be 10-25\% in the range $4 < \pt < 12 \ \gevc$. 
The $\pt$ spectrum and nuclear modification factor for $\jpsi$ with $\pt < 10 \ \gevc$ is reported, along with results from $\Upsilon$(1S+2S+3S) production in Au+Au collisions. 
The nuclear modification factor for high-$\pt$ $\jpsi$ ($\pt > 5 \ \gevc)$ is found to be consistent with unity in peripheral collisions, 
while a significant suppression of low-$\pt$ $\jpsi$ and $\Upsilon$ is observed in central collisions. 
The elliptic flow of $\jpsi$ is reported for 20-60\% central Au+Au collisions, and is found to be consistent with zero.

}

 \FullConference{XXIst International Europhysics Conference on High Energy Physics\\
 21-27 July 2011\\
 Grenoble, Rhones Alpes France}

\input{Commands.tex}

\begin{document}

 %
 \section{Introduction}
 %


The suppression of $\jpsi$ and other heavy quarkonium is expected due to the Debye color-screening of the potential between heavy quarks in a dense medium, 
and has been suggested as a signature of the formation of Quark Gluon Plasma (QGP)~\cite{ref:matsui}. 
A suppression of $\jpsi$ production in heavy ion collisions has been observed by the NA50 and NA60 experiments 
at the CERN-SPS~\cite{ref:na50}~\cite{ref:na60}. 
A similar amount of $\jpsi$ suppression was observed at RHIC at midrapidity in $\snn = 200 \ \gev$ Au+Au collisions~\cite{ref:phenix}, despite
the increased collision energy as compared to SPS. 
This can be understood by considering other effects due to the presence of a QGP which may contribute to the modification of heavy quarkonium production, 
such as statistical coalescence of heavy quark-antiquark pairs or co-mover absorption~\cite{ref:recombination, ref:recombination2}. There are also contributions to the inclusive yields from 
feed-down effects~\cite{ref:feeddown} and the sequential melting of excited $\jpsi$ and $\Upsilon$ states~\cite{ref:melting}.
Cold Nuclear Matter (CNM) effects~\cite{ref:cnm}, such as the modification of nuclear PDFs (shadowing)~\cite{ref:shadowing}, 
and final state nuclear absorption~\cite{ref:absorption}, need to be taken into account in order to fully quantify an anomalous suppression. 
This can be achieved by studying the production of various quarkonium states in $\pp$, d+A and A+A collisions. 
Furthermore, $\pp$ collisions can offer insight to the quarkonium production mechanism, as no model can yet fully explain the 
observed $\jpsi$ yields in elementary collisions. 
The $\jpsi$ elliptic flow ($v_{2}$) can provide information about the contribution from recombination of charm quarks and the degree of thermalization of charm quarks in the medium. 
\\

The results for the $\jpsi$ $\pt$ spectrum in $\pp$ and Au+Au collisions at $\snn = 200 \ \gev$ recorded by the STAR detector in 2009 and 2010 are presented in this paper. The centrality 
dependence of the $\jpsi$ and $\Upsilon$(1S+2S+3S) nuclear modification factor ($\raa$) are also shown. 
The elliptic flow of $\jpsi$ is reported for 20\--60\% central Au+Au collisions at $\snn = 200 \ \gev$.

 %
 %

 %
 \section{Results}

 The reconstruction of $\jpsi$ and $\Upsilon$ are done via the dielectron decay channel, $\jpsi, \Upsilon \rightarrow e^{+} + e^{-}$, with a branching ratio (B) of 5.9\% for $\jpsi$ and 
 2.5\%, 1.9\%, and 2.2\% for the 1S, 2S, and 3S $\Upsilon$ states, respectively. 
 The reconstruction methods used in this analysis are similar to those used in the 2005 and 2006 data~\cite{ref:run6pp, ref:run6ppups}, 
 and are described in~\cite{ref:run10AuAuDetail, ref:run10AuAu, ref:run10ups}. 
 The high-$\pt$ $\jpsi$ results shown in this paper are from 2009 $p+p$ collisions with an integrated luminosity of $1.8 pb^{-1}$, and 2010 Au+Au 
 collisions with an integrated luminosity of $1.4 nb^{-1}$. The data were collected using the Barrel Electromagnetic Calorimeter (BEMC) 
  to trigger events with a single tower ($\Delta  \eta \times \Delta \phi = 0.05 \times 0.05$) above the transverse energy threshold $E_{T} > 2.6 \ \gev$ in $p+p$ 
  collisions and  $E_{T} > 4.3 \ \gev$ in Au+Au collisions. 
 Over 300M minimum bias and 250M central-biased Au+Au events were also recorded, which allow for a precise low-$\pt$ ($\pt < 5 \ \gevc$) $\jpsi$ analysis. 
 The Time Of Flight (TOF) detector, which covers full azimuth and $|\eta| < 0.9$, was installed in 2009 with 72\% completed for recording $\pp$ data, and 
 was fully installed in 2010 for the Au+Au data recording. The Time Projection Chamber (TPC), which is responsible for momentum and $\dedx$ reconstruction, has been used 
 along with the TOF and BEMC to obtain a $\jpsi$ signal with increased signal-to-background ratio and statistical precision.\\

\begin{figure}[h]
\begin{center}$
\begin{array}{cc}
\includegraphics[width=.46\textwidth, height=55mm]{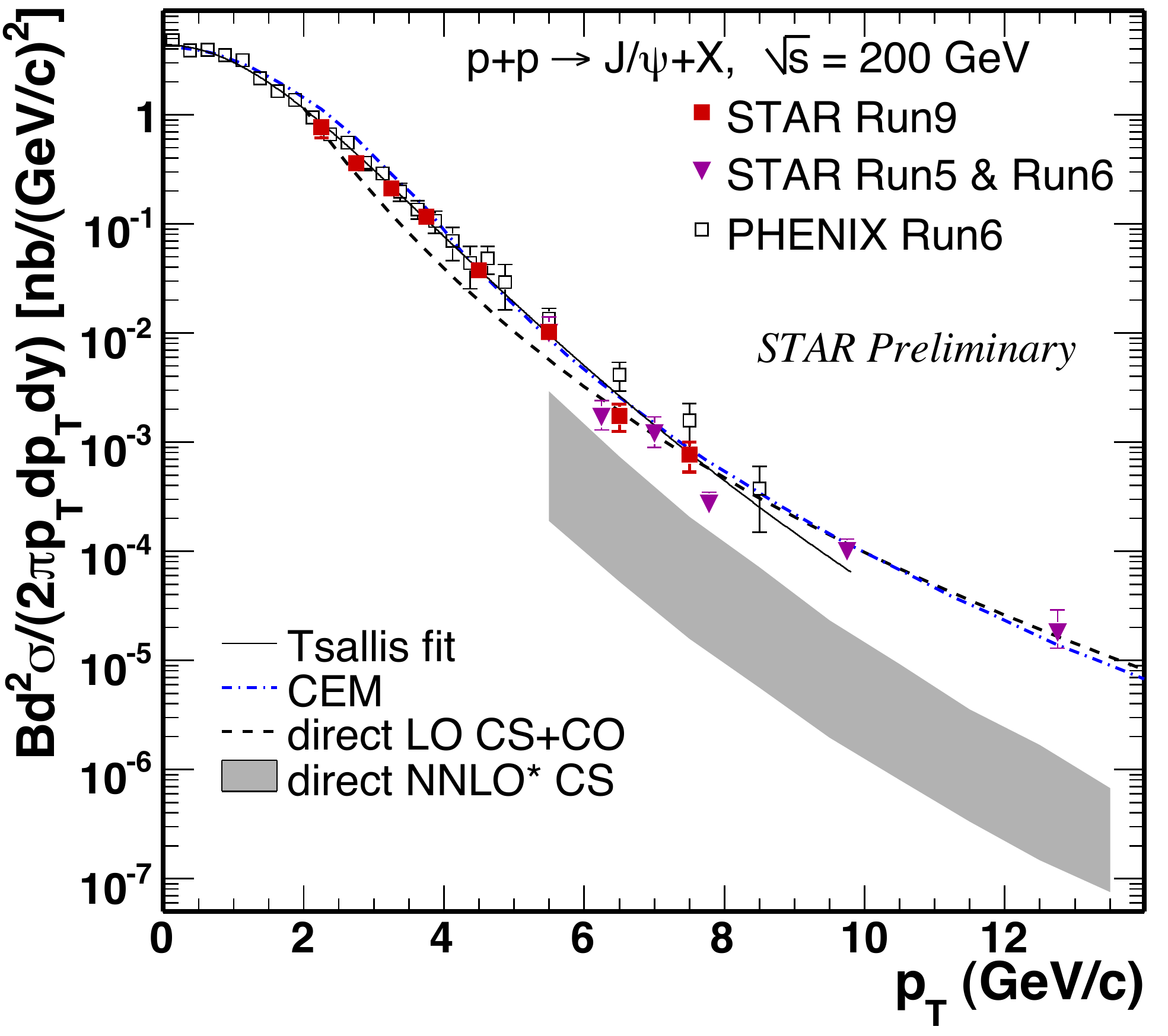} &
\includegraphics[angle=90, width=.46\textwidth, height=55mm]{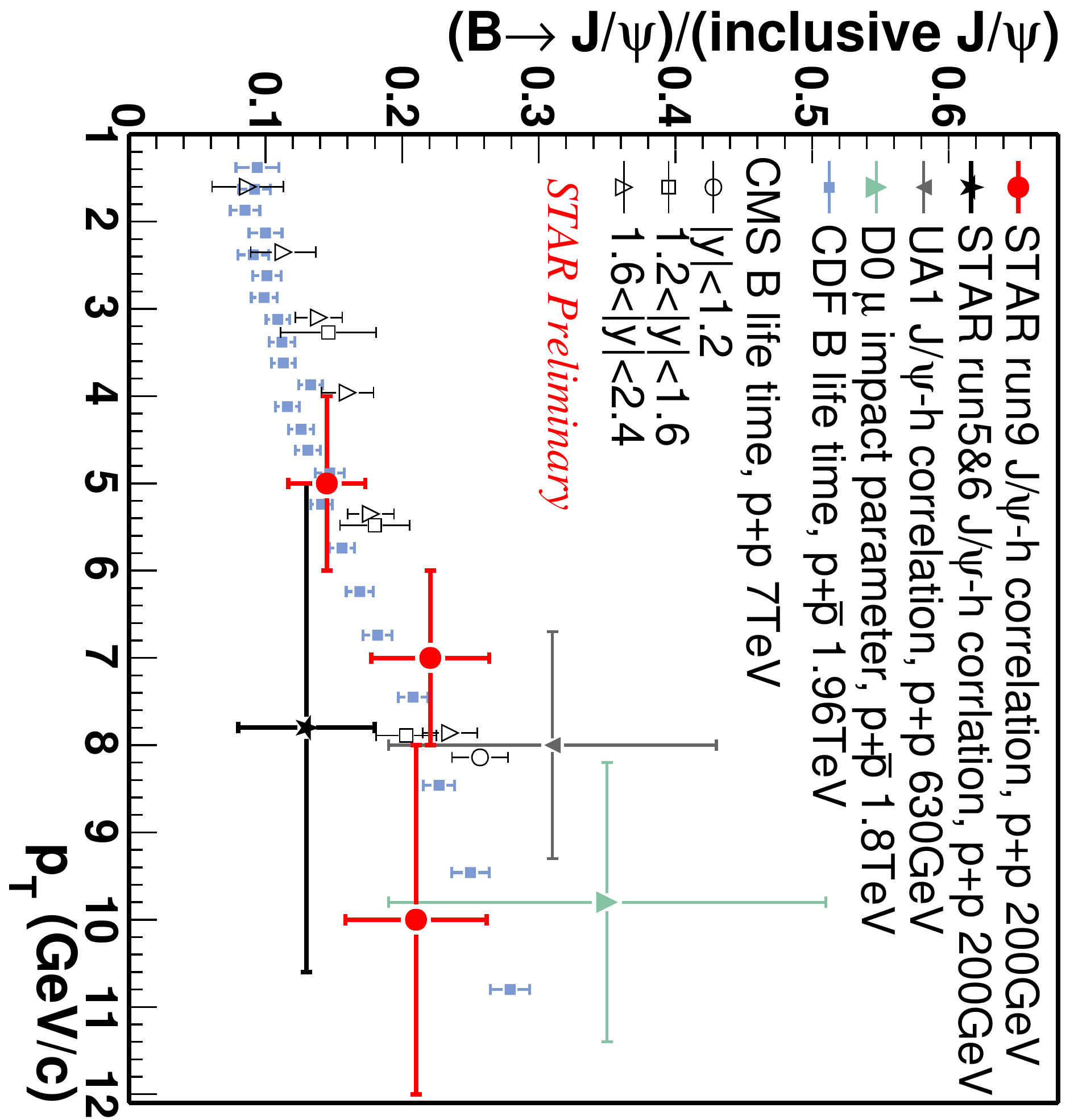} \\
\end{array}$
\caption{ The $\jpsi$ $\pt$ spectra (left) and $B$-meson feed-down contribution (right) in $\pp$ collisions.
}
\label{fig:pp}
\end{center}
\end{figure}

The left panel of Fig.~\ref{fig:pp} shows the corrected $\pt$ spectrum for $\jpsi$ in $\pp$ collisions at $\snn = 200 \ \gev$ (closed square). 
This is consistent with the previous STAR (downward triangle)~\cite{ref:run6pp} and PHENIX (open square)~\cite{ref:phenixpp} measurements. 
A Blast-Wave model based on Tsallis statistics (TBW)~\cite{ref:tbw1, ref:tbw2} has been fitted to all data points (solid line). 
Predictions from NRQCD using LO color-singlet (CS) and color-octet (CO)~\cite{ref:co} 
transitions are shown (dashed line), and agree with the data reasonably well but leave no room for feed-down effects 
from $B$, $\chi_{c}$, and $\psi'$, which may account for up to 50\% on the inclusive yield. 
The predictions from NNLO$^{*}$ CS~\cite{ref:cs} (grey band) underestimate the yield and $\pt$ shape at high-$\pt$. 
The Color Evaporation Model (CEM) has been shown (dot-dashed line), and can explain the data reasonably well~\cite{ref:cem}. The right panel of Fig.~\ref{fig:pp} shows the $\pt$ dependence 
of the $B \rightarrow \jpsi$ feed-down contribution (closed circle). These results are extracted from $\jpsi$-hadron azimuthal angular correlations using PYTHIA, and 
indicate a feed-down contribution of 10-25\%, which is similar to measurements at higher collision energies~\cite{ref:feed1, ref:feed2, ref:feed3, ref:feed4, ref:feed5, ref:feed6}.\\ 

\begin{figure}[ht!!!!!!!!!!!!!!!!!!!!!]
\begin{center}$
\begin{array}{cc}
\includegraphics[width=.46\textwidth, height=55mm]{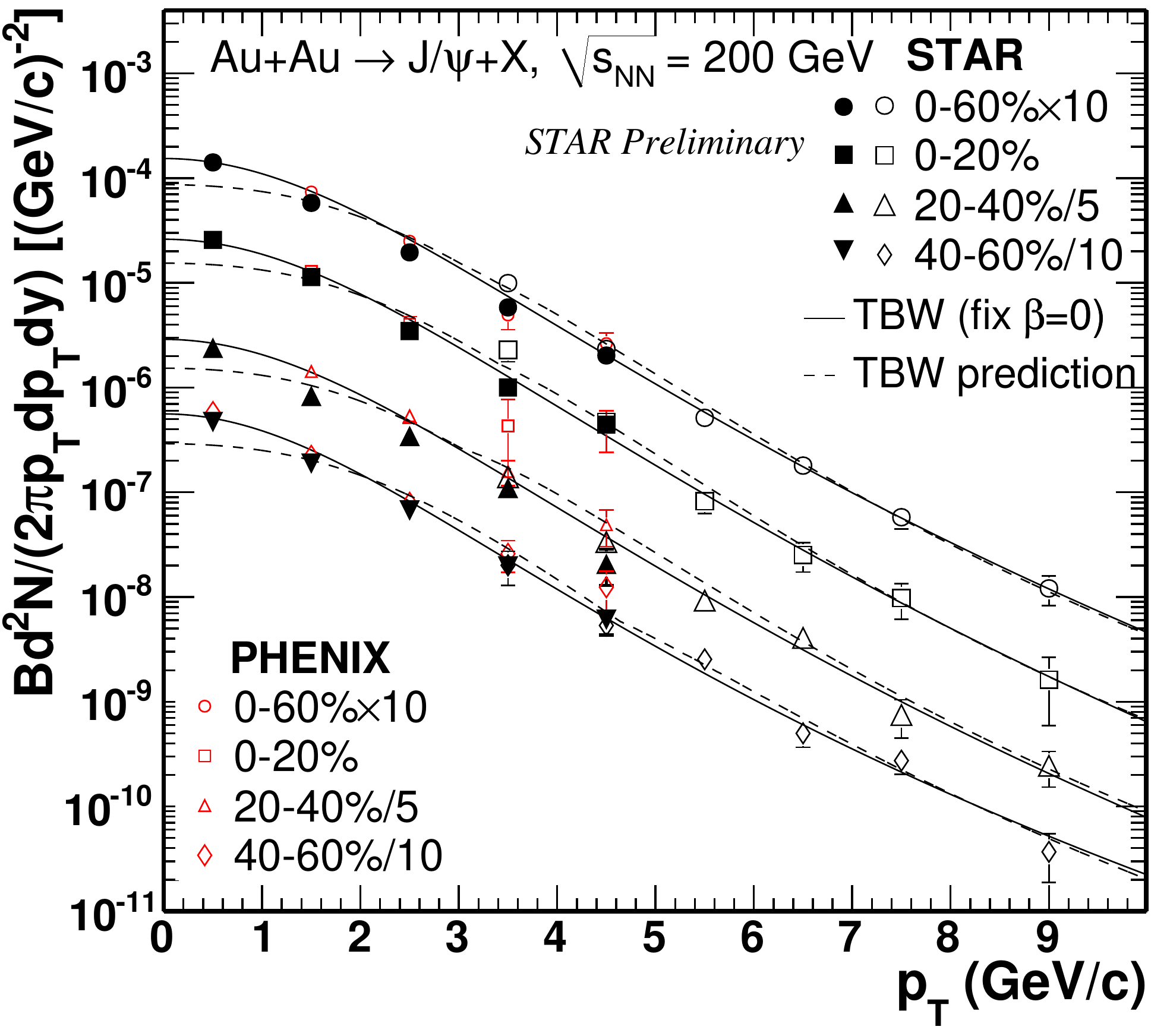} &
\includegraphics[width=.46\textwidth, height=55mm]{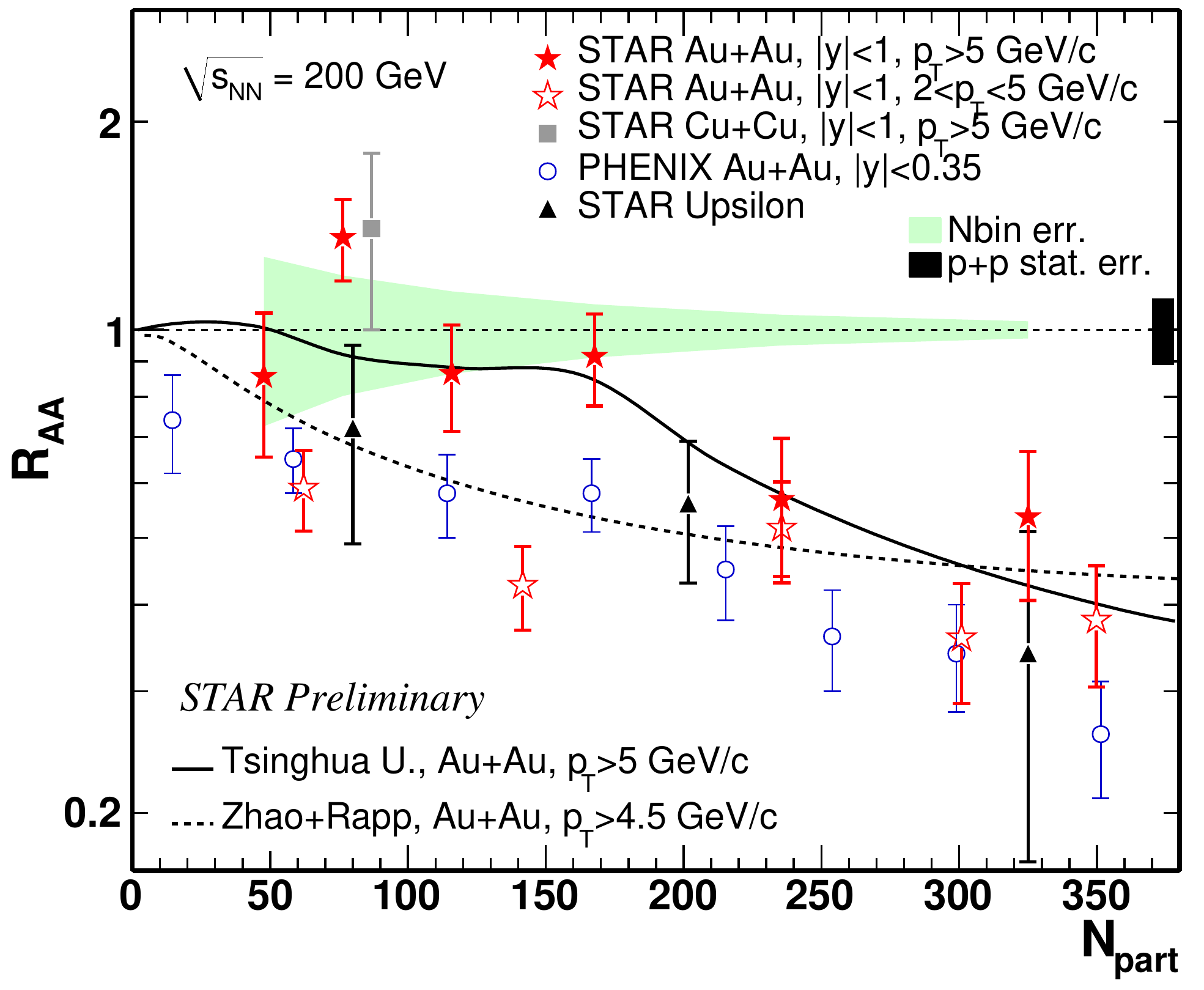} \\
\end{array}$
\caption{ The $\pt$ spectra for $\jpsi$ (left) and nuclear modification factor for $\jpsi$ and $\Upsilon$(1S+2S+3S) (right) in Au+Au collisions.
}
\label{fig:auau}
\end{center}
\end{figure}

The corrected $\pt$ spectrum for low-$\pt$ $\jpsi$ (solid black symbols) and high-$\pt$ (open black symbols) are 
shown in Fig.~\ref{fig:auau} (left panel), and are consistent with the previously published data (open red symbol)~\cite{ref:phenix}. 
A TBW prediction using the freeze-out properties from light hadrons (dashed line) is compared to the data, and agrees with the data for $\pt > 2 \ \gevc$. 
The agreement is improved when fixing the radial flow ($\beta$) to zero. This may suggest that the $\jpsi$ has a smaller radial flow than light hadrons, or that regeneration 
from charm quarks may be a significant contribution at low-$\pt$.\\

The right panel of Fig.~\ref{fig:auau} shows the nuclear modification factor of $\jpsi$ and $\Upsilon$(1S+2S+3S) as a function of the number 
of participant nucleons ($\npart$) in Au+Au collisions at $\snn = 200 \ \gev$. 
The low-$\pt$ ($2 < \pt < 5 \ \gevc$) $\jpsi$ (open star) and $\Upsilon$ (upward triangle) show a suppression that 
increases with increasing collision centrality, with $\raa =0.3-0.4$ 
in central collisions. These results are consistent with previous measurements (open circle)~\cite{ref:phenix}. The high-$\pt$ ($\pt > 5 \ \gevc$) $\jpsi$ $\raa$ is systematically above the 
low-$\pt$ measurements for all centralities, and is consistent with unity in peripheral collisions (20\--60\%). This is consistent with previous high-$\pt$ Cu+Cu measurements made 
by STAR (closed square)~\cite{ref:run6pp}.\\ 

\begin{figure}[ht!!!!!!!!!]
\begin{center}
\includegraphics[width=.55\textwidth]{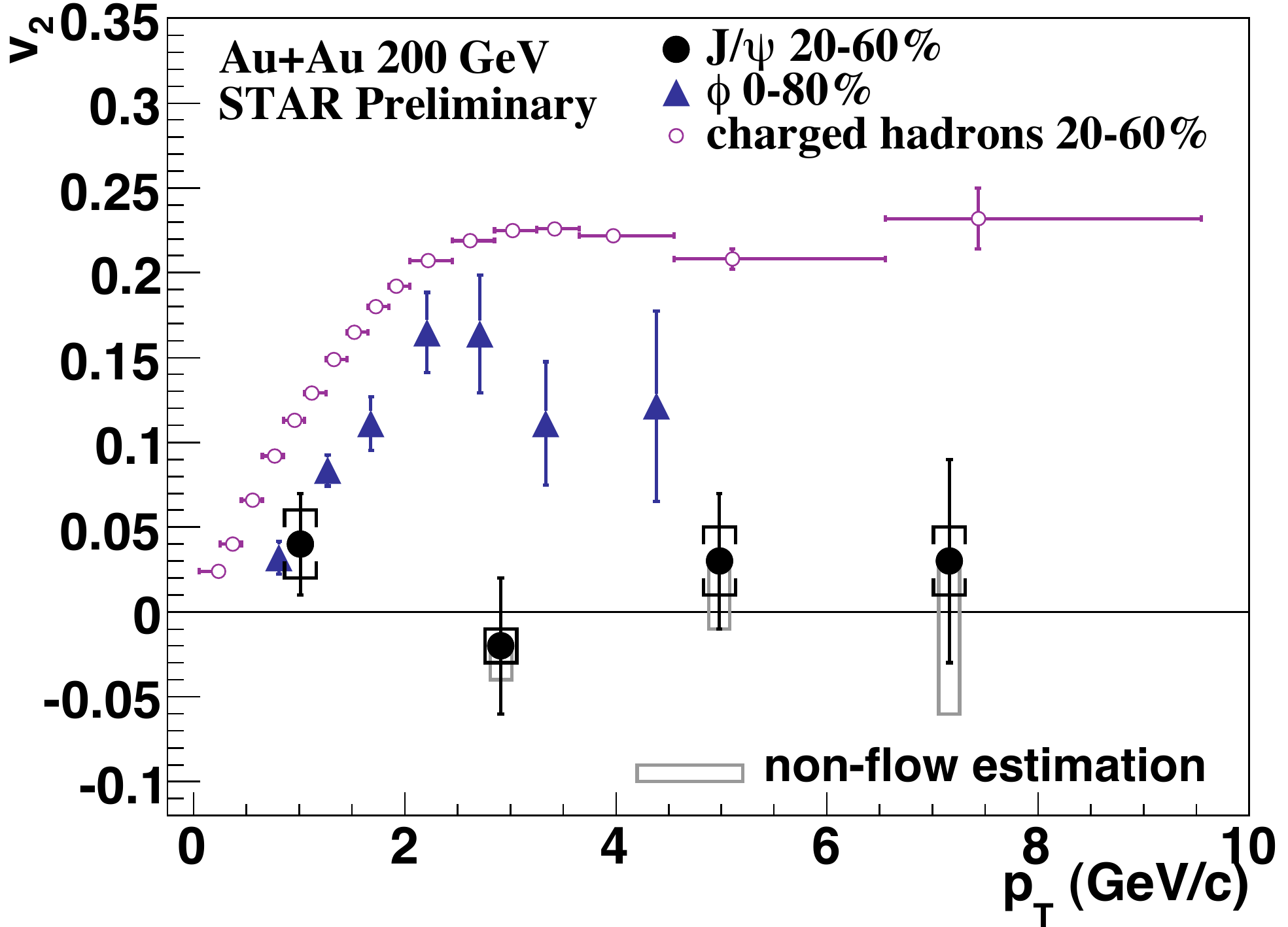} \\
\caption{ The $\jpsi$ elliptic flow, $v_{2}$ for 20\--60\% central collisions. 
}
\label{fig:v2}
\end{center}
\end{figure}

Primordial $\jpsi$s which are produced in the initial hard scattering are expected to carry very little flow, however those which are subsequently created from the statistical coalescence 
of charm quarks may carry non-zero elliptic flow. 
The $\jpsi$ elliptic flow is shown in Fig.~\ref{fig:v2} for mid-central Au+Au collisions at $\snn = 200 \ \gevc$ (closed circle), and is consistent with zero within errors. 
This is compared to the $v_{2}$ from charged hadrons (open circle) and $\phi$-mesons (upward triangle), which exhibit a non-zero elliptic flow. The small $v_{2}$ of $\jpsi$ in 
mid-central collisions may indicate that the complete thermalization of charm quarks is not achieved, or that recombination of charm quarks is a small contribution in this kinematic range. \\

 %
 %


\end{document}

%% file: Commands.tex
\newcommand{\jpsi}{\textrm{J}/\psi}

\newcommand{\pp}{\textit{p}+\textit{p}}

\newcommand{\snn}{\mbox{$\sqrt{s_{_{NN}}}$}}

\newcommand{\npart}{\mbox{$N_{\textrm{part}}$}}

\newcommand{\raa}{\mbox{$R_{\textit{AA}}$}}

\newcommand{\dedx}{\mbox{$\mathrm{d}E/\mathrm{d}x$}}

\newcommand{\pt}{\mbox{$p_T$}}

\newcommand{\gev}{\mbox{$\mathrm{GeV}$}}
\newcommand{\gevc}{\mbox{$\mathrm{GeV/}c$}}

